# Integration of Quantum Key Distribution in a 20-km 32-user Coherent Passive Optical Network with Single Feeder Fiber

**Jing Wang, Brian J. Rollick, Zhensheng Jia, Bernardo A. Huberman**
*CableLabs, 858 Coal Creek Circle, Louisville, CO, 80027, USA*
j.wang@cablelabs.com

**Abstract:** For the first time, we demonstrated the integration of O-band polarization-encoding decoy-state BB84 QKD into a C-band 20-km single-feeder fiber 32-user coherent PON running at carrier-grade power levels without modifying existing PON infrastructures. 

## 1. Introduction

Quantum key distribution (QKD) provides an information-theoretic secure way to establish secret keys between two communication parties [1]. Most developed QKD protocols, however, require dedicated fibers and point-to-point (P2P) links and are not optimized for cost-sensitive applications [2]. On top of this, delivering secure keys to end users remains a challenge since the last few miles are covered by point-to-multipoint (P2MP) passive optical networks (PONs), where fiber scarcity and cost limitations make it difficult to add or dedicate any fibers for QKD. Several techniques have been reported to integrate QKD into PONs. Dual-feeder fibers were first proposed to integrate QKD in WDM-PONs [3] with the additional feeder fiber dedicated to QKD to avoid Spontaneous Raman scattering (SpRS) noise from classical data. Then the same method was also applied to GPON [4-6], 10G-EPON [5], and NG-PON2 [6]. Another method is to reduce the power of classical data. For GPONs with single feeder fiber, the power of classical data must be attenuated significantly to allow normal operation of QKD; otherwise, the feeder fiber has to be limited to 1 km [5]. Although Ref [7] claimed the first integration of QKD in a single-feeder GPON, two groups of optical network units (ONUs) are connected to separate splitters and two feeder fibers. The upstream (US) data was also attenuated with only 4 km feeder fiber. Other methods include bypassing the splitter by creating a wavelength detour for the QKD channel [8] or embedding QKD pulses into 0 bits of classical data to avoid SpRS noise in the time domain [9]. So far, most works require modifications to either fiber networks (e.g., adding feeder fiber, bypassing splitter) or transceivers (synchronization with QKD pulses) in PONs, facing challenges in terms of scalability and cost. Integrating QKD into PONs without changing existing PON infrastructure has remained an unresolved topic.

Since current PONs exploit S, C, and O-bands for classical upstream and downstream (DS) data, the broadband SpRS noise contaminates all O, S, C, and L bands. Also, intensity-modulation direct-detection (IM-DD) schemes exploited by current PONs feature low receiver sensitivity and high launch power, leading to even worse SpRS noise. Luckily, the ever-growing demand for user capacity pushed coherent optics into PONs [10] with both US/DS data in the C-band to leverage coherent ecosystems from long-haul and metro applications. As a result, the O-band remains available for QKD operation. Thanks to the large wavelength separation, SpRS noise from C to O-band is significantly reduced. Moreover, coherent optics' high receiver sensitivity and low launch power provide additional benefits to minimize Raman noise. By placing the QKD channel in the O-band and classical data in the C-band, for the first time, we propose integrating QKD into coherent PONs without altering existing PON infrastructures [11]. In this work, the coexistence of an O-band polarization-encoding decoy-state BB84 QKD channel with C-band 100-Gb/s dual-polarization (DP)-QPSK US/DS data is demonstrated in a 20-km 32-user PON with single-feeder fiber and carrier-grade power levels up to 9.2 dBm, making this a promising solution to deploy QKD in brown-field PONs.

## 2. Operation Principles

Fig. 1(a) shows the architecture for integrating QKD into coherent PONs. Both DS/US data channels exploit coherent optics in the C-band, leaving QKD in the O-band. The QKD channel is (de)multiplexed with classical data by coarse WDM (CWDM) multiplexers. Since a QKD transmitter (QTx) consists of commercial off-the-shelf devices, whereas a QKD receiver (QRx) relies on expensive and delicate single-photon detectors (SPDs), to lower the overall cost, one set of SPDs is deployed at the optical line terminal (OLT) and shared by many optical network units (ONUs). Only one set of QTx is needed to add a new user, and no upgrade is required at the OLT. In this upstream QKD scenario, QKD pulses co-propagate with US data and counter-propagate with DS data. The SpRS noise collected by SPDs consists of backscattering noise from DS data and forward-scattering noise from US data, with the dominant noise source coming from backscattering in the feeder fiber because the DS data in the feeder fiber has the highest power and noise photons backscattered in the feeder fiber do not have to pass through the splitter to reach SPDs. For forward-scattering noise from the US data, either US data or noise must pass through the splitter before reaching SPDs.

## 3. Experimental Setup

Fig. 1(b) shows the experimental setup. We built a polarization-encoding decoy-state BB84 QKD system at 1310 nm. At the QTx in ONU, an intensity modulator (IM) generates pulses and prepares decoy states with a pulse width of 200 ps and a repetition rate of 25 MHz. The polarization modulator (Pol-M) consists of a circulator, two polarization beam splitters (PBSs), one phase modulator (PM), and a Faraday mirror (FM) [12]. $PBS_1$ separates two orthogonal polarization states into two arms, one modulated by the PM, and then recombined by $PBS_2$. The FM reflects the recombined light with polarization rotated by 90˚. The two orthogonal polarizations are separated by $PBS_2$ and pass through opposite arms on their way back. By adjusting the voltage on PM, different phase shifts are induced between two polarizations to prepare four polarization states in two conjugate bases, 0˚, 90˚, 45˚, and -45˚. The variable optical attenuator ($VOA_1$) adjusts the pulse intensity leaving ONU at point B to single-photon levels. A three-intensity decoy-state protocol was used with average photon numbers per pulse of 0.8, 0.04, and $10^{-3}$, and emission probabilities of 0.9, 0.05, and 0.05 for signal, decoy, and vacuum states, respectively. A 0.213 nm narrowband filter at 1310 nm was used to eliminate out-of-band noise before the QRx at OLT. A 4 ns window was used to gate four SPDs to eliminate the out-of-window noise. The SPDs have 15% detection efficiency, 10 µs dead time, and a dark count of $2.4\times10^{-6}$ per gate. The optical misalignment of the system is 1.5-2%. Both DS/US data use 100 Gb/s DP-QPSK channels with wavelengths tunable across the C-band. They are (de)multiplexed with QKD by the CWDM multiplexer at OLT/ONU. Since the SpRS noise is dominated by the backscattering from DS data in the feeder fiber, an EDFA is used to boost the launch power of DS data for the worst-case scenario. Up to 9.2 dBm DS launch power (at point A) is tested, which is one order of magnitude higher than carrier-grade power levels in conventional PONs. The QKD synchronization channel is omitted since a 25 MHz optical clock has too low power to impact QKD performance.

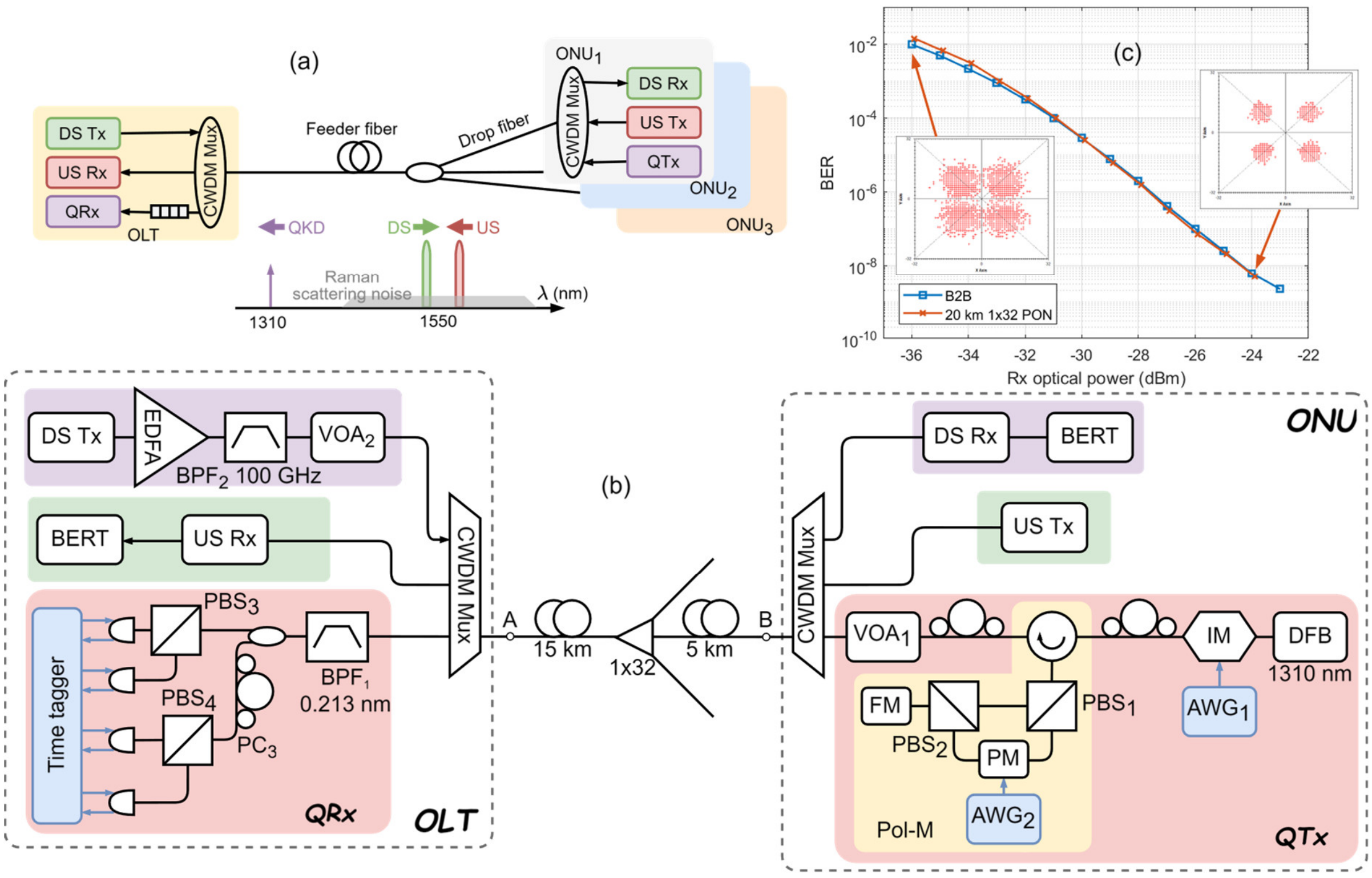

Fig. 1. (a) Integration of an O-band QKD channel into a C-band coherent PON. (b) Experimental setup. (c) BER performance of 100-Gb/s DP-QPSK links for B2B and after 20 km 32-user PON.

## 4. Experimental Results

Fig. 1(c) shows the bit error rate (BER) performance of classical 100 Gb/s DP-QPKS channels vs received optical power (identical for both DS/US channels) with QPSK constellations in the insets. Both back-to-back (B2B) and 20 km 32-user PON cases are shown, with negligible power penalty thanks to the high receiver sensitivity of the coherent receiver. Fig. 2(a) shows the backscattering noise as a function of the launch power of DS data, measured at point A in Fig. 1(b). The dark count of SPDs without DS data is plotted by the dashed line as a reference. The backscattering noise increases with DS power. Several DS wavelengths across the C-band are tested. Longer wavelengths generate less noise because they are further away from the 1310 nm QKD channel. Fig. 2(b) shows the forward-scattering noise from US data vs the launch power of US data, measured at point B in Fig. 1(b). The forward-scattering noise stays constant around the dark count level for US power up to 3.2 dBm. Various US wavelengths were tested across the C-band, all with negligible forward-scattering noise. Fig. 2(c, d) shows the quantum BER (QBER) and secure key rates (SKR) as functions of the launch power of DS data with the US channel turned off. Consistent with the noise measurement in Fig. 2(a), QBER increases and SKR decreases with the DS launch power, and longer DS wavelength leads to better performance due to the smaller SpRS noise. At 0 dBm, SKR of at least $7\times10^{-6}$ bits per pulse (175 b/s) is achieved for all DS wavelengths in the C-band. For DS power up to 3.2 dBm, secure keys can be generated for all DS wavelengths in the C-band. For DS power up to 9.2 dBm, QKD operation is still possible for DS wavelengths at and longer than Ch13 (191.3 THz, 1567.13 nm). This suggests that O-band QKD and L-band DS channels would be a promising combination for QKD integration in coherent PON. Fig. 2(e, f) shows the QBER and SKR with an active US channel of 3.2 dBm launch power. Consistent with Fig. 2(b), forward-scattering noise from US data has a negligible impact on QKD. So, the performance in Fig. 2(e, f) is identical to Fig. 2(c, d). For key rate estimation, we followed the asymptotic limit of infinite key size, since the QKD link can operate continuously.

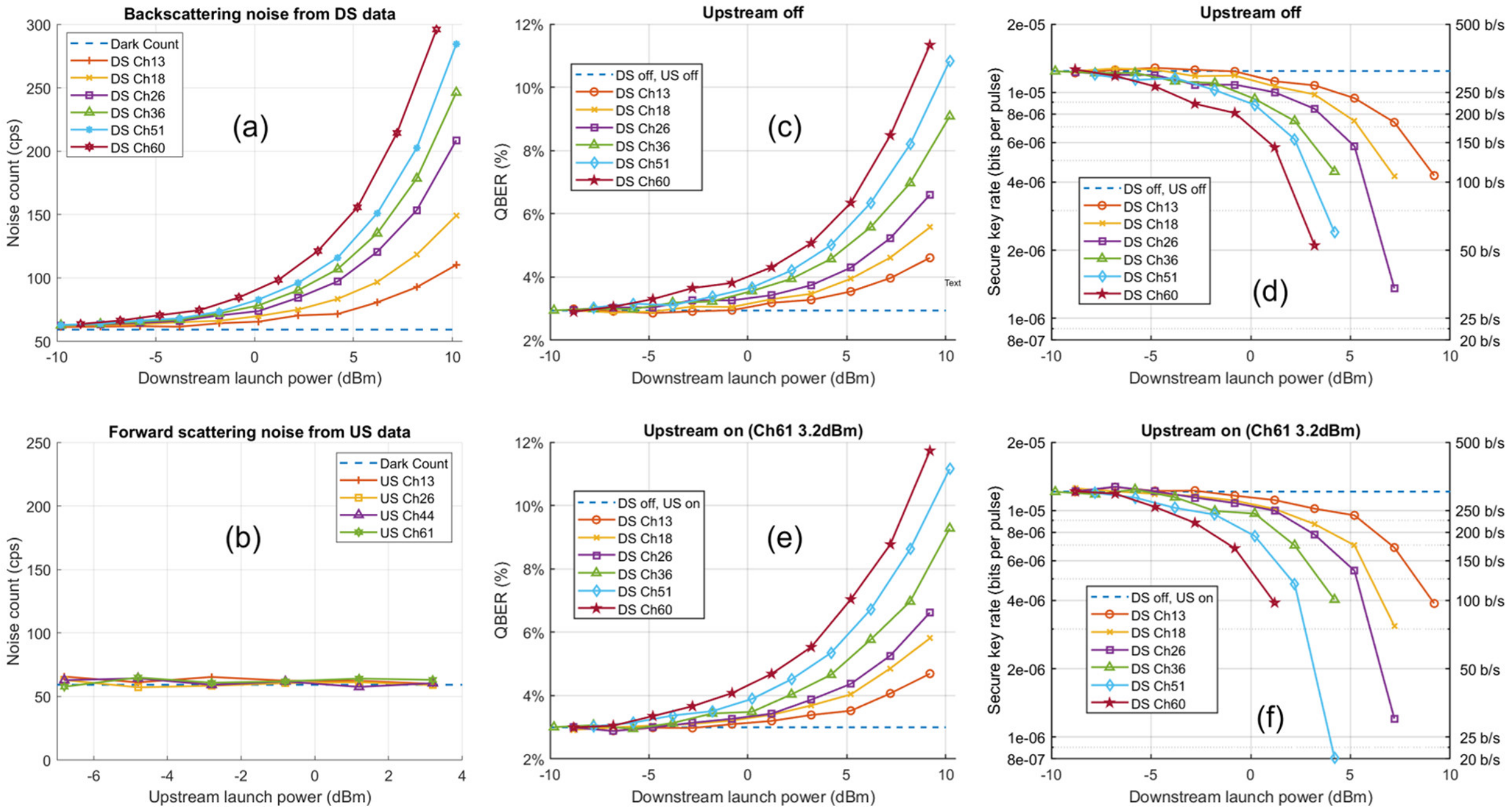

Fig. 2. Experimental results. (a) Backscattering noise from DS data as a function of the launch power of DS data. (b) Forward-scattering noise from US data as a function of the launch power of US data. (c, d) QBER and SKR vs DS launch power without US. (e, f) QBER and SKR vs DS launch power with a 3.2 dBm US channel.

## 5. Conclusions

For the first time, we demonstrated the integration of QKD in coherent PONs without modifying existing PON infrastructure. The coexistence of an O-band polarization-encoding decoy-state BB84 QKD channel and C-band 100-Gb/s DP-QPSK US/DS data is demonstrated in a 20-km 32-user coherent PON with single feeder fiber. QKD operation is allowed with classical communication at carrier-grade power levels up to 9.2 dBm.